\documentclass[showpacs,preprintnumbers,amsmath,amssymb,aps,twocolumn,superscriptaddress,letterpaper,nofootinbib]{revtex4-1}

\usepackage{amsfonts}

\usepackage{epsfig,amssymb,euscript}
\usepackage{type1cm}
\usepackage{color}

\newcommand{\reef}[1]{(\ref{#1})}

\def\be{\begin{equation}}
\def\ee{\end{equation}}
\def\bea{\begin{eqnarray}}
\def\eea{\end{eqnarray}}

\begin{document}
\title{Holographic Schwinger effect and the geometry of entanglement}
\author{Julian Sonner}
\email{sonner@mit.edu}
\affiliation{Center for Theoretical Physics, Massachusetts Institute of Technology\\ Cambridge, MA 02139, U.S.A.}
\affiliation{L.N.S., Massachusetts Institute of Technology\\ Cambridge, MA 02139, U.S.A.}

\preprint{MIT-CTP 4483}


\begin{abstract} 
We show that the recently proposed bulk dual of an entangled pair of a quark and an anti-quark corresponds to the Lorentzian continuation of the tunneling instanton describing Schwinger pair creation in the dual field theory. This observation supports and further explains the claim by Jensen \& Karch \cite{Jensen:2013ora} that the bulk dual of an EPR pair is a string with a wormhole on its world sheet. We suggest that this constitutes an AdS/CFT realization of the creation of a Wheeler wormhole.
\end{abstract}


\maketitle

\section{Introduction}
\begin{figure*}[t!]
\begin{center}
a)\includegraphics[width=0.20\textwidth]{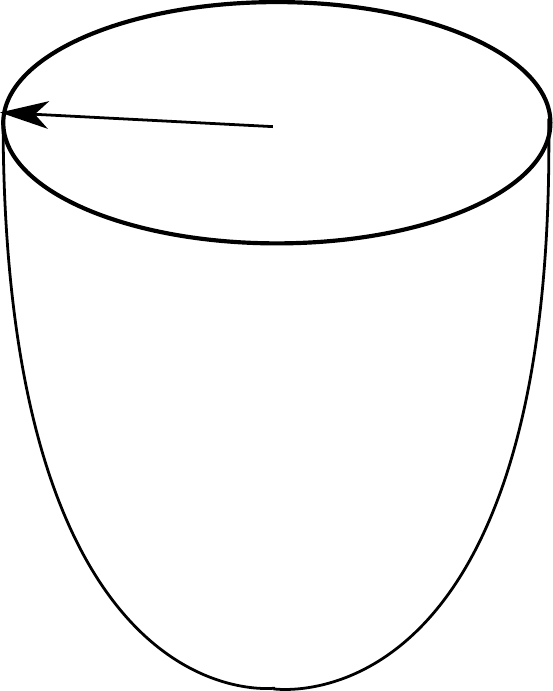}\qquad\qquad\qquad b)\includegraphics[width=0.3\textwidth]{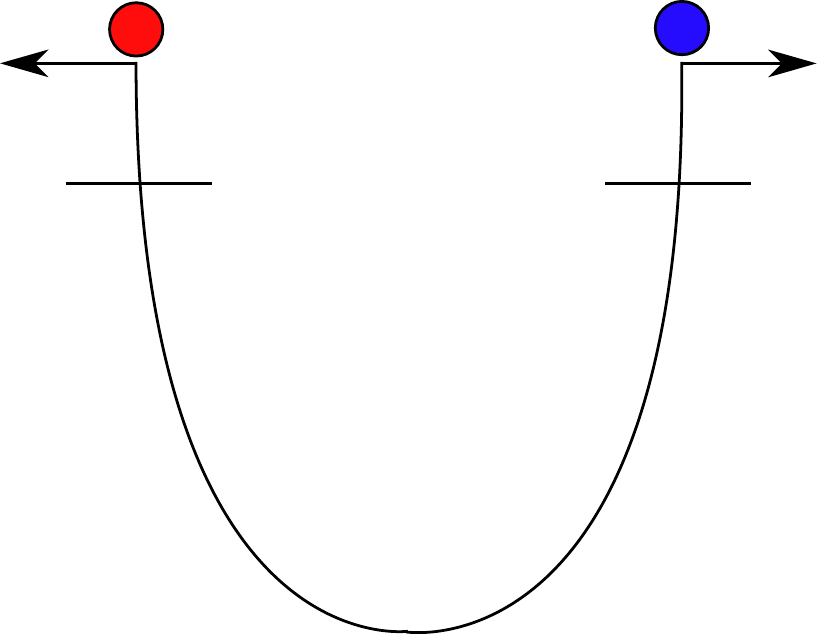}
\vskip1.5em
 \begin{picture}(0.1,0.25)(0,0)
\put(-9.6,4.2){\makebox(120,100){${\rm quark}$}}
\put(-9.6,4.2){\makebox(50,100){${\rm anti-quark}$}}
\put(-9.6,4.2){\makebox(150,68){{\small $\frac{1}{r}=\sqrt{R^2 + \frac{1}{r_0^2}}$}}}
\put(-9.6,4.2){\makebox(-60,75){$R$}}
\put(-9.6,4.2){\makebox(-50,0){${\rm tunneling}\quad{\rm instanton}$}}
\put(-9.6,4.2){\makebox(80,0){${\rm EPR}\quad{\rm pair}$}}
\end{picture}
 \caption{\it The Euclidean solution describing Schwinger pair production is an instanton ending at $r=r_0$ in a circle of radius $R$. Its Lorentzian continuation is the EPR geometry corresponding to an entangled pair of (anti-) quarks with two world sheet horizons at $\frac{1}{r}=\sqrt{R^2 + \frac{1}{r_0^2}}$ and a wormhole connecting them. The instanton for $n>1$ is the $n-$cover of the instanton for $n=1$ and gives the contribution to the production probability due to $n$ pairs. The figure has been analytically continued to Lorentzian time from the left to the right panel and the circle $\tau^2 + x^2$ has been transformed in to the hyperbola $-t^2 + x^2$.\label{fig.DiscInstantonLorentz}}
\end{center}\end{figure*}
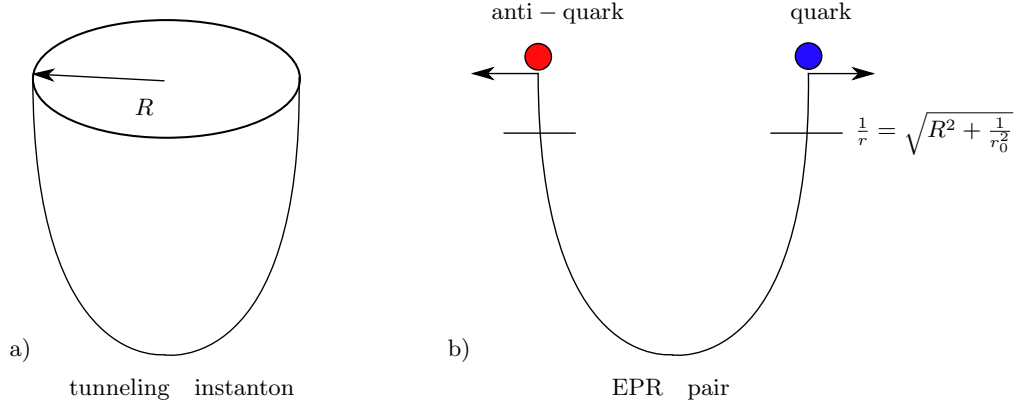

Maldacena and Susskind, sparked by the firewall debate \cite{Almheiri:2012rt,Braunstein:2009my}, conjectured that two seemingly completely disparate physical phenomena, the entanglement of quantum states on one side, and non-traversable wormholes on the other, are in fact intimately related \cite{Maldacena:2013xja}. This conjecture has been summarised in an equation between acronyms\footnote{Here and throughout the Letter ER = Einstein-Rosen \cite{PhysRev.48.73} and EPR = Einstein-Podolsky-Rosen \cite{einstein1935can}.}, ER = EPR. While the authors of \cite{Maldacena:2013xja} have formulated their conjecture for general entangled pairs, there has recently emerged a specific context in which the ER = EPR conjecture can be made very concrete. It was argued in \cite{Jensen:2013ora} that - within the AdS/CFT duality - an EPR pair made from a quark and an antiquark has a bulk dual described by a string whose world-sheet has a non-traversible wormhole of the kind the authors of \cite{Maldacena:2013xja} were considering. In this Letter we show that the exact geometry put forward as the requisite bulk dual of an EPR pair in \cite{Jensen:2013ora} arises as the Lorentzian continuation of a Euclidean string instanton that describes the production of entangled pairs via the Schwinger effect in the strongly coupled dual gauge theory (see Fig. \ref{fig.DiscInstantonLorentz}).

This observation puts the interpretation of the solution in \cite{Jensen:2013ora} in terms of entanglement beyond reasonable doubt and strongly supports the ER = EPR conjecture in the context of the  ${\cal N}=4$ SYM theory at large $N$ and large 't Hooft coupling $\lambda$  and its gravity dual. We cannot resist but remark that the picture proposed here is rather reminiscent of the `Wheeler wormhole' idea \cite{wheeler1964geometrodynamics} in that the particle anti-particle pair in a maximally entangled singlet state is connected by a wormhole, created together with the pair itself as a consequence of the Schwinger effect. Of course the wormhole in this context lives on a string world sheet whose metric is merely induced from the surrounding gravity theory, while the more general ER = EPR conjecture is still open (see also \cite{Nikolic:2013dva,Marolf:2013dba,Chowdhury:2013mka}). We hope that the observations made in this Letter may be of help in exploring the issue further.
\section{Creating the entangled pair of \cite{Jensen:2013ora} with the Schwinger effect at strong coupling}
The Schwinger effect is a phenomenon that occurs in quantum field theory in the presence of a strong applied field. The prototypical context, within which Julian Schwinger first derived the eponymous effect \cite{Schwinger1951}, gives the probability of pair production of charged particles within a space-time volume $V$, of mass $m$ in an applied field $E$, as $P=1-e^{-\gamma V}$, with
\be\label{eq.QEDSchwinger}
\gamma = \frac{ E^2}{8 \pi^3}\sum_{n=1}^\infty \frac{(-1)^{n+1}}{n^2}e^{-\frac{\pi m^2 n}{|E|}}\,,
\ee
for particles of spin zero\footnote{The result for spin $j$ is also known - see for example the description in \cite{Semenoff:2011ng}. }.
Since the particles so produced are formed from a singlet state (the vacuum) they are necessarily entangled with one another, no matter what the actual nature of the particles, may they be electrons and positrons as in the original  case, or quarks and anti-quarks and even charged W bosons, as we shall consider in this Letter. The exponential factor in Eq. \reef{eq.QEDSchwinger} strongly suggests a derivation of the effect in terms of an instanton sum, where the multi-instanton contributions are suppressed at weak fields and/or large masses, and we now describe very briefly an approach in which this is made precise.

An enlightening treatment of the Schwinger effect is obtained by considering the world-line path integral of a particle in Euclidean signature \cite{Affleck:1981bma,Dunne:2005sx,Dunne:2006st,Semenoff:2011ng}. In this formalism the pair creation effect can be derived by considering the saddle-points of the Euclidean path integral, which are given by cyclotron orbits of the particle, with the $n$-instanton contribution given by a particle going around the orbit $n$ times along the trajectory $\hat r = \left(\cos(2\pi n \tau), \sin(2\pi n \tau),0,0\right)$. Recall that we have cyclotron orbits because an electric field acts like a magnetic field once we continue to Euclidean signature. The interpretation of the $n$-covered circle is the creation of $n$ pairs instead of just one \cite{lebedev1984virial,Schwinger1951,Dunne:2005sx}. By evaluating the fluctuation determinants around the classical saddle point configuration and summing over all $n$ one arrives at \reef{eq.QEDSchwinger}.

Semenoff and Zarembo \cite{Semenoff:2011ng}, building on earlier work by \cite{Berenstein:1998ij, Drukker:1999zq,Gorsky:2001up}, have used this point of view to give a strong-coupling derivation of the Schwinger effect in ${\cal N}=4$ SYM theory\footnote{See also the earlier paper by \cite{Gorsky:2001up} which first applies the instanton worldsheet utilised by \cite{Semenoff:2011ng} to the Schwinger effect.} on its Coulomb branch, where the original gauge group is broken from $U(N+1)$ to $U(1)\times U(N)$. We refer the reader to the original references for details. The crux of their computation is that at strong coupling, where the ${\cal N}=4$ theory is described by strings in $AdS_5\times S^5$, one can similarly calculate the rate of Schwinger pair production by finding a suitable instanton configuration, in their case for a string world sheet embedded in $AdS_5\times S^5$ ending on a D3 brane a distance $r_0$ away from the Poincar\'e horizon. This gives the rate of pair production of massive $W^\pm$ bosons in the Coulomb branch of the theory. We can reinterpret their computation in terms of a pair-creation process of a quark and an anti-quark in a $U(N)$ gauge theory with fundamental quarks. Now the string ends on a flavor brane at $r=r_0$ and one can apply an electric field with respect to the $U(1)_B$ baryon number symmetry of the theory. Alternatively we could stick with the original interpretation of pair-produced $W$ bosons, and phrase the argument of \cite{Jensen:2013ora} in terms of such entangled states of $W^\pm$ bosons.  To be specific, let us take the $AdS_5\times S^5$ metric as
\be
ds^2 = L^2 \left(  r^2 dx_\mu dx^\mu + \frac{dr^2}{r^2} + d\Omega_5^2\right)\,.
\ee
Then, as shown in \cite{Semenoff:2011ng} the relevant configuration giving the $n$-instanton contribution is a world-sheet, parametrized by $\tau,\sigma$, whose Euclidean embedding coordinates satisfy the equation\footnote{This corrects a trivial typo in Eq. (15) of \cite{Semenoff:2011ng}. I thank Gordon Semenoff for corespondence on this issue.}
\be
X = \frac{\cosh(2\pi n \sigma_0)}{\cosh(2\pi n \sigma)}R \hat r\,,\qquad r = r_0 \frac{\tanh (2\pi n \sigma_0)}{\tanh(2\pi n \sigma)}\,.
\ee
Here $\hat r$ parametrizes the circular trajectory introduced above and $\sinh{(2\pi n \sigma_0)}=1/(R r_0)$, while $n$ is the analog of the number of orbits of the particle we saw above. The radius $R$ is obtained by extremizing the instanton action with the result \cite{Semenoff:2011ng}
\be
R = \frac{1}{2\pi m} \sqrt{\left(\frac{2\pi m^2}{E}\right)^2 - \lambda}\,,
\ee
in terms of the mass $m = \frac{\sqrt{\lambda}r_0}{2\pi}$ and applied field $E$. We show a representation of the instanton in Fig. \ref{fig.DiscInstantonLorentz}a.
Continuing this world-sheet configuration to Lorentzian signature, one finds that it becomes the locus of the hyperbola
\be\label{eq.lorentzEmbedding}
-t^2 + x^2 + \frac{1}{r^2} = R^2 + \frac{1}{r_0^2}\,,
\ee
evidently a solution corresponding to a quark and an anti-quark (or a pair of W$^\pm$ bosons) being accelerated away from each other and approaching the speed of light in asymptotic time \cite{Xiao:2008nr}, as illustrated in Fig. \ref{fig.DiscInstantonLorentz}b. In analogy with the (scalar) QED case, one is tempted to conclude that the solution for $n>1$ corresponds to the instanton saddle relevant for the creation of $n$ pairs, but it is not clear that the standard argument used there \cite{lebedev1984virial}, applies to strongly-coupled non-Abelian theories. What is clear is that this process contributes to the production rate with a suppression $\propto e^{-n S_1}$, where $S_1$ is the one-instanton result (cf. Eq. \reef{eq.QEDSchwinger}).

\begin{figure}[h!]
\begin{center}
\includegraphics[width=0.5\textwidth]{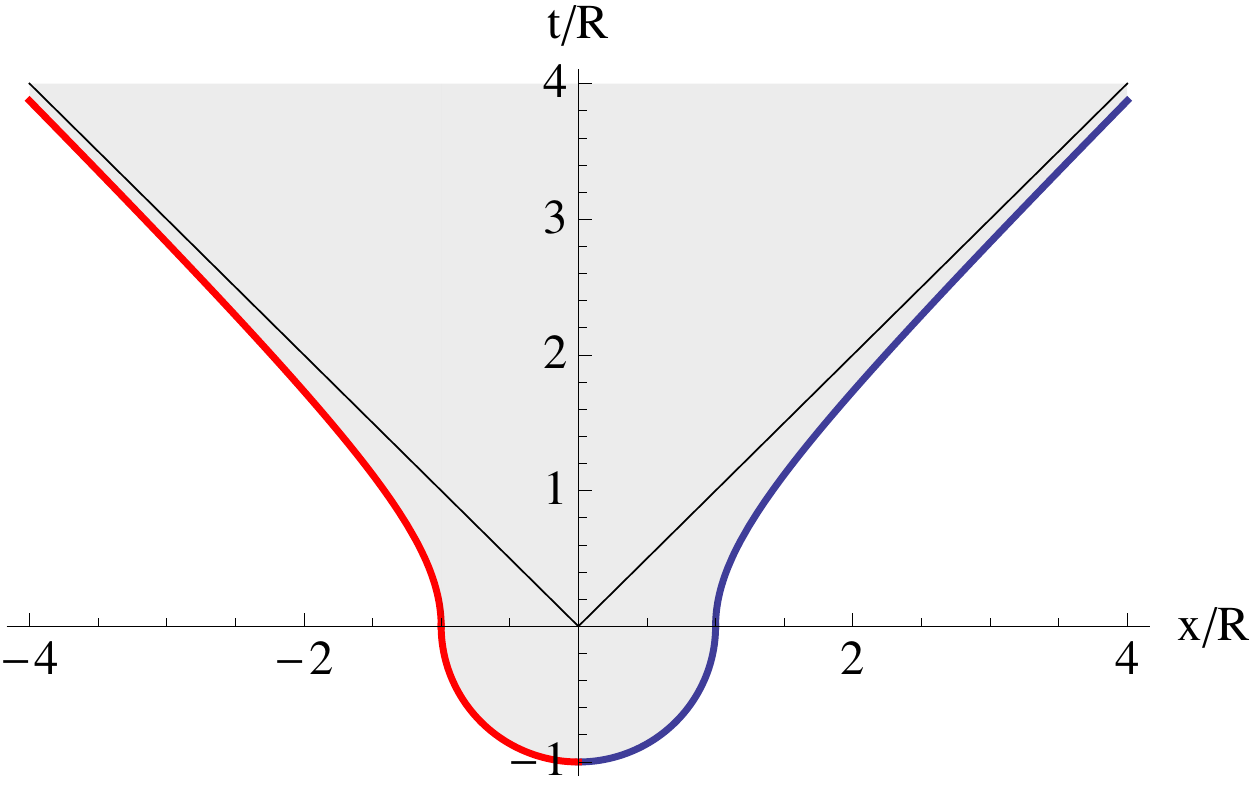}
 \begin{picture}(0.1,0.25)(0,0)
\put(-9.6,4.2){\makebox(50,10){${\rm Euclidean}$}}
\put(-9.6,4.2){\makebox(80,30){${\rm Lorentzian}$}}
\end{picture}
 \caption{\it  In the upper half plane the end points of the string follow the hyperbola \reef{eq.lorentzEmbedding} and the world sheet horizon is induced from the Rindler horizon of the pair shown as solid black lines. The world sheet of the string extends in the direction perpendicular to the $x-t$ plane, filling in the space between the blue and red curves (shaded in gray). The Rindler horizons project to the world sheet horizons at $r= \left(R^2 + 1/r_0^2  \right)^{-1/2}$. The full solution arises by analytically continuing the Euclidean instanton through $t=0$, which we depict by drawing the instanton-circle trajectory in the lower half plane. The causal structure is thus the same as the upper half of the Penrose diagram of the eternal AdS black hole and inherits its ER bridge.  \label{fig.CausalStruct}}
\end{center}\end{figure}
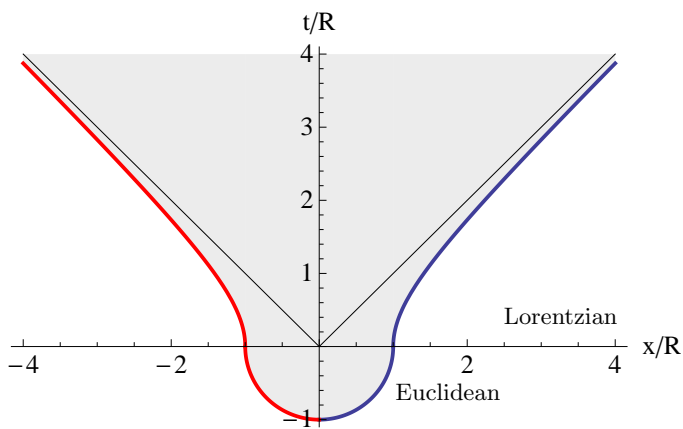

We are now in a position to state the main argument of this Letter: The analytically continued  world-sheet instanton is precisely the configuration put forward by \cite{Jensen:2013ora} as a solution for the dual of an entangled pair of quarks in ${\cal N}=4$ SYM. In order to compare to their solution we simply have to make the identification $b^2 = R^2 + \frac{1}{r_0^2}$ and $z = 1/r$. As the authors of \cite{Jensen:2013ora} pointed out, the world-sheet configuration has brane horizons at $z=b$ connected by a non-traversible Lorentzian wormhole. The causal structure of the world sheet corresponding to pair production is illustrated in Fig. \ref{fig.CausalStruct} (see caption for further explanation).

 In this Letter we added the new point that the configuration can be viewed as arising from the tunneling instanton describing Schwinger pair creation in the field theory under consideration. This makes it obvious that the state is entangled and strongly supports the claim that ER = EPR for maximally entangled states in the context of gauge-gravity duality. Obviously here we should read the equals sign as `is dual to'. Furthermore we would like to identify the Lorentzian embedding \reef{eq.lorentzEmbedding} as the `Wheeler wormhole' created together with the pair.   It would be very interesting to consider extensions to states with less than maximal entanglement.

\section{Discussion}
This Letter has focused on the Schwinger process in strongly coupled ${\cal N}=4$ SYM in $d=3+1$, but brane horizons, double-sided or not, have appeared in many other contexts, for example \cite{Chernicoff:2008sa,Caceres:2010rm,Xiao:2008nr,Das:2010yw,Gursoy:2010aa,Gubser:2006nz,Kim:2011qh,Alam:2012fw}.
We expect that similar arguments can be made about any brane configuration with a two-sided horizon. Furthermore, in all cases we are aware of, (stationary) brane horizons that are not induced on the world-sheet - or world volume - by an actual black-hole horizon of the background, but rather appear due to some external forcing, correspond to driven non-equilibrium (steady) states \cite{Das:2010yw,Sonner:2012if}. The entropy associated with such horizons, derived by assuming that the world-sheet/volume geometry can be taken literally, should be associated with the entropy of entanglement of the Schwinger pairs being continually produced or similar entangling (pair-creation) effects. Indeed, it was already pointed out in \cite{Karch:2010kt,Sonner:2012if}, that the world-sheet horizon of the D3/D5 system in the presence of a strong applied field is associated with the Schwinger effect. Since the brane horizon depends on the applied field one can dial its area at will by changing the field intensity up and down. Thus the brane horizon cannot satisfy the second law of BH thermodynamics in a conventional sense and so it is satisfying that we are nevertheless able to put forward an explanation for the observed entropy, namely that it represents the entropy of entanglement of the pairs being produced in the driven steady state. This rate of production goes to zero when the external forcing is switched off, which also removes the brane horizon.

It is often stated (see for example \cite{Mathur:2012np}) that particle creation at a black-hole horizon is subtly different from the conventional Schwinger effect and also that entanglement entropy is not enough in order to understand the horizon entropy of general black holes. In this work, we have argued that the two can be dual to each other in a precise sense. We add the cautionary note that the metric whose horizon we a ssociate with Schwinger pair production is not that of dynamical gravity in itself, but rather an induced one; and indeed it has been observed before that the black-hole entropy in induced gravity theories can be entirely accounted for by a suitable entanglement entropy, as reviewed for example in \cite{Jacobson:1994iw}.

We conclude by restating the main point: The solution of  \cite{Jensen:2013ora}, previously found in \cite{Xiao:2008nr}, is the Lorentzian continuation of a family of world-sheet instantons that describes particle creation via the Schwinger effect in ${\cal N}=4$ SYM theory. Therefore the interpretation of this solution as the bulk dual of a maximally entangled pair is correct and the AdS dual of an EPR pair is a geometry with an ER bridge. This is very reminiscent of an old idea due to Wheeler. At least at strong 't Hooft coupling and large $N$ we have been able to explicitly demonstrate that a particle and an anti-particle, pair produced in an applied field, are connected by a wormhole (in the dual geometry), and so this wormhole should be associated with the entanglement between them. It is tempting to ask what becomes of this wormhole at weak 't Hooft coupling $\lambda$, where stringy corrections need to be taken into account in the bulk. More daunting is the task of elucidating the fate of the ER bridge under $1/N$ corrections when the dual gravity theory ceases to be classical and the object connecting the particle and anti-particle becomes very quantum mechanical. At least in principle, the framework described in this Letter offers a starting point to look into these interesting questions.
\subsection*{Acknowledgements}
I would like to thank Mike Crossley, Hong Liu, Josephine Suh and Wojciech Zurek for discussions and comments on the draft version, and Gordon Semenoff for helpful correspondence. I thank Los Alamos National Laboratory and the Asia Pacific Center for Theoretical Physics (APCTP) for hospitality during the course of this work. This work was supported in part by the U.S. Department of Energy (DOE) under cooperative research agreement Contract Number DE-FG02-05ER41360.

\bibliographystyle{apsrev}

\bibliography{SchwingerTwoColumnPublished}{}

\end{document}